\documentclass[12pt]{article}
\usepackage{graphicx,graphics}
\usepackage{amsmath}
\usepackage{amsfonts}
\usepackage{amssymb}
\title{Some exact solutions for the Caudrey-Dodd-Gibbon equation}

\author{Alvaro Salas\thanks{Department of Mathematics, Universidad de
Caldas, Department of Mathematics, Universidad Nacional de Colombia,
Manizales, Colombia. \emph{email} : asalash2002@yahoo.com}}
\date{}
\begin {document}
\maketitle
\begin {abstract}
In this paper we show some exact solutions for the
Caudrey-Dodd-Gibbon equation (CDG equation). These solutions are
obtained via \circledR \emph{Mathematica} 6.0 by the projective
Riccati equation method.
\end{abstract}
\section{Introduction}
From 70's, a vast variety of the simple and direct methods to find
analytic solutions of nonlinear differential equations and evolution
equations have been developed. Recently, the projective Riccati
equation method has drawn lots of interests in seeking the solitary
wave solution and other kinds of solutions.\newline
 The general form of
the fifth-order KdV equation is written as
\begin{equation}\label{eq0}
u_t+\omega\, u_{xxxxx}+\alpha\, uu_{xxx}+\beta\, u_xu_{xx}+\gamma\,
u^2u_x=0
\end{equation}
where $\alpha$, $\beta$, $\gamma$ and $\omega$ are arbitrary real
parameters. In this paper we present exact solutions for a
particular case of this equation obtained by setting $\alpha=30$,
$\beta=30$, $\gamma=180$ and $\omega=1$ . Thus, the equation to be
studied is
\begin{equation}\label{eq1}
u_t+ u_{xxxxx}+ 30uu_{xxx}+30u_xu_{xx}+180u^2u_x=0
\end{equation}
Equation (\ref{eq1}) is known as  Caudrey-Dodd-Gibbon equation.
\section{The projective Riccati equation method}
We search exact solutions of equation (\ref{eq1}) in the form
\begin{equation}\label{eq2}
\begin{cases}
u(x,t)=v(\xi)\\
\xi=x+\lambda t,
\end{cases}
\end{equation}
As a result we have that the equation (\ref{eq1}) is reduced to the
nonlinear ordinary differential equation (ODE)
\begin{equation}\label{eq3}
180 V'(\xi ) V(\xi )^2+30 V^{(3)}(\xi ) V(\xi
   )+\lambda  V'(\xi )+30 V'(\xi ) V''(\xi
   )+V^{(5)}(\xi )=0
\end{equation}

To obtain exact solution for the equation (\ref{eq3}), we use the
projective Riccati equation method [4][7] which may be described in
the following three steps:\\\\
\noindent \textit{\textbf{Step 1}}.  We consider solutions of
(\ref{eq3}) in the form
\begin{equation}\label{eq4}
v(\xi)=a_0+\sum_{j=1}^m
\sigma(\xi)^{j-1}(a_j\sigma(\xi)+b_j\tau(\xi)),
\end{equation}
where $\sigma(\xi)$, $\tau(\xi)$  satisfy the system
\begin{equation}\label{eq5}
\begin{cases}
\sigma'(\xi)=e\sigma(\xi)\tau(\xi)\\
\tau'(\xi)=e\tau^2(\xi)-\mu \sigma(\xi)+r.
\end{cases}
\end{equation}
It may be proved that the first integral of this system is given by
\begin{equation}\label{eq6}
\tau^2=-e\left[r-2 \mu
\sigma(\xi)+\dfrac{\mu^2+\rho}{r}\sigma^2(\xi)\right],
\end{equation}
where $\rho=\pm 1$ and $e=\pm 1$. \\\noindent We consider the
following solutions of the system (\ref{eq5}).
\begin{enumerate}
\item \textbf{Case I}:
\newline
If $r=\mu =0$ then
\begin{equation}
\tau_1(\xi)=-\frac{1}{e\xi},\;\;\sigma_1(\xi)=\frac{C}{\xi}.
\end{equation}
\item \textbf{Case II}:
\newline
If $e=1$, $\rho=-1$ and $r>0$ :
\begin{equation}
\begin{cases}
\sigma_1(\xi )=
    \dfrac{r\,\sec ({\sqrt{r}}\,\xi )}
  {1 + \mu \,\sec ({\sqrt{r}}\,\xi )};\quad
\tau_1(\xi )=
   \dfrac{{\sqrt{r}}\,\tan ({\sqrt{r}}\,\xi )}
  {1 + \mu \,\sec ({\sqrt{r}}\,\xi )}\\\\
\sigma_2(\xi )=\dfrac{r\,\csc ({\sqrt{r}}\,\xi )}
  {1 + \mu \,\csc ({\sqrt{r}}\,\xi )};\quad
\tau_2(\xi )=-\dfrac{{\sqrt{r}}\,\cot ({\sqrt{r}}\,\xi )}
  {1 + \mu \,\csc ({\sqrt{r}}\,\xi )}.
\end{cases}
\end{equation}
\item \textbf{Case III}:
\newline
If $e=-1$, $\rho=-1$ and $r>0$ :
\begin{equation}
\begin{cases}
\sigma_3= \dfrac{r\,\text{sech}({\sqrt{r}}\,\xi )}
  {1 + \mu \,\text{sech}({\sqrt{r}}\,\xi )}\\\\
\tau_3=\dfrac{{\sqrt{r}}\,\tanh ({\sqrt{r}}\,\xi )}
  {1 + \mu \,\text{sech}({\sqrt{r}}\,\xi )}.
\end{cases}
\end{equation}
\item \textbf{Case IV}:
\newline
If $e=-1$, $\rho=1$ and $r>0$ :
\begin{equation}
\begin{cases}
\sigma_4= \dfrac{r\,\text{csch}({\sqrt{r}}\,\xi )}
  {1 + \mu \,\text{csch}({\sqrt{r}}\,\xi )}\\\\
\tau_4=\dfrac{{\sqrt{r}}\,\coth ({\sqrt{r}}\,\xi )}
  {1 + \mu \,\text{csch}({\sqrt{r}}\,\xi )}.
\end{cases}
\end{equation}
\end{enumerate}
$$$$
\noindent\textit{\textbf{Step 2}}. Substituting (\ref{eq4}), along
with (\ref{eq5}) and (\ref{eq6}) into (\ref{eq3}) and collecting all
terms with the same power in $\sigma^{i}(\xi)\tau^{j}(\xi)$,  we get
a polynomial in the two variables $\sigma(\xi)$ and $\tau(\xi)$.
This polynomial has the form
\begin{eqnarray}
a\sigma(\xi)^{m+5}+b\sigma(\xi)^{2m+3}+c\sigma(\xi)^{m+4}\tau(\xi)+d\sigma(\xi)^{3m+1}+\nonumber\\
+e\sigma(\xi)^{2m+2}\tau(\xi)+\text{other terms of lower
degree}\label{eq7}
\end{eqnarray}
We assume that $m\geq 1$ to avoid trivial solutions.  The degrees of
the highest terms are $m+5$ ( the degree of the terms
$a\sigma(\xi)^{m+5}$ and $c\sigma(\xi)^{m+4}\tau(\xi)$), $2m+3$ (
the degree of the term $b\sigma(\xi)^{2m+3}$ ) and $3m+1$ ( the
degree of the term $d\sigma(\xi)^{3m+1}$ ). There are two integer
values of $m$ for which $3m+2=2m+3$ or $3m+1=m+5$ or $2m+3=m+5$.
These are $m=1$ and $m=2$. We are going to find solutions for $m=1$
( the case $m=2$ is not considered here ).  When $m=1$  solutions
have the form $v(\xi)=a_0+ a_1\sigma(\xi)+b_1\tau(\xi)$ and equating
in (\ref{eq7}) the coefficients of every power of $\sigma(\xi)$ and
of every term of the form $\sigma^j(\xi)\tau(\xi)$ to zero, we
obtain the following algebraic system in the variables $a_0$, $a_1$,
$b_1,\ldots$ : \newline\newline\noindent
$\begin{array}{ll}
 \textbf{1.} & e^7 \left(\mu ^2+\rho
   \right)^2 a_1=0
   \end{array}$\newline
$\begin{array}{ll}
  \textbf{2.} & e^8 \left(\mu ^2+\rho
   \right)^3 b_1=0 \\
   \end{array}$\newline
$\begin{array}{ll}
 \textbf{3.} & e^5 \left(\mu ^2+\rho
   \right)^2 \left(6 \mu
   e^3-3 \mu  e+4 a_1\right)
   b_1=0 \\
  \end{array}$
  \newline
$\begin{array}{ll}
 \textbf{4.}& e^4 \left(\mu ^2+\rho
   \right) \left(2 r \mu  a_1
   e^3-\mu ^2 b_1^2 e-\rho
   b_1^2 e-r \mu  a_1 e+r
   a_1^2\right)=0  \end{array}$
 \newline
$\begin{array}{ll}
  \textbf{5.} & (e-1) (e+1) r \left(2 r
   e^3-r e-3 a_0\right)
   b_1^2=0
    \end{array}$
 \newline
$\begin{array}{ll}
 \textbf{6.} & 120 r^2 a_1 e^7-180 r^2
   a_1 e^5-180 r a_0 a_1
   e^4+61 r^2 a_1 e^3+150 r
   a_0 a_1 e^2+180 a_0^2 a_1
   e+\lambda  a_1 e \\
  & -960 r \mu  b_1^2 e^5+1230
   r \mu  b_1^2 e^3+720 \mu
   a_0 b_1^2 e^2-540 r a_1
   b_1^2 e^2-300 r \mu  b_1^2
   e-360 \mu  a_0 b_1^2\\
   &+360 r
   a_1 b_1^2=0  \end{array}$
 \newline
$\begin{array}{ll}
 \textbf{7.} & 16 r^2 \mu  a_1 e^7-20
   r^2 \mu  a_1 e^5+8 r^2
   a_1^2 e^4-12 r \mu  a_0 a_1
   e^4+5 r^2 \mu  a_1 e^3-6
   r^2 a_1^2 e^2+6 r \mu  a_0
   a_1 e^2 \\
  & -48 r \mu ^2 b_1^2 e^5-16
   r \rho  b_1^2 e^5+46 r \mu
   ^2 b_1^2 e^3+12 r \rho
   b_1^2 e^3+12 \mu ^2 a_0
   b_1^2 e^2-12 r a_0 a_1^2
   e-6 r \mu ^2 b_1^2 e \\
  &+12 e^2 \rho  a_0 b_1^2-36
   e^2 r \mu  a_1 b_1^2+12 r
   \mu  a_1 b_1^2=0  \end{array}$
 \newline
$\begin{array}{ll}
 \textbf{8.} & 24 r \mu ^2 a_1 e^6+8 r
   \rho  a_1 e^6-22 r \mu ^2
   a_1 e^4-6 r \rho  a_1 e^4-6
   \mu ^2 a_0 a_1 e^3-6 \rho
   a_0 a_1 e^3+3 r \mu ^2 a_1
   e^2 \\
  & -32 \mu ^3 b_1^2 e^4-32
   \mu  \rho  b_1^2 e^4+16 r
   \mu  a_1^2 e^3+17 \mu ^3
   b_1^2 e^2+17 \mu  \rho
   b_1^2 e^2-7 r \mu  a_1^2
   e+6 r a_1^3\\
     & -18 e \mu ^2 a_1 b_1^2-18
   e \rho  a_1 b_1^2=0 \\
\end{array}$ \newline
$\begin{array}{ll} \textbf{9.} & (e-1) (e+1) r b_1
  (120 r^2 e^6-120 r^2
   e^4-180 r a_0 e^3+16 r^2
   e^2-180 r b_1^2 e\\
   &+60 r a_0
   e+180 a_0^2+\lambda
   )=0 \end{array}$ \newline
$\begin{array}{ll} \textbf{10.} & b_1(720 r^2 \mu  e^8-1320
   r^2 \mu  e^6+662 r^2 \mu
   e^4-61 r^2 \mu  e^2+2
   \lambda  \mu  e^2-150 r \mu
    a_0 e-\lambda  \mu  \\
  & -720 r \mu  a_0 e^5+480
   r^2 a_1 e^5+840 r \mu  a_0
   e^3-540 r^2 a_1 e^3+360 \mu
    a_0^2 e^2+90 r^2 a_1 e-180
   \mu  a_0^2 \\
  & -720 r \mu  b_1^2 e^3-720
   r a_0 a_1 e^2+540 r \mu
   b_1^2 e+360 r a_0 a_1)=0 \end{array}$ \newline
$\begin{array}{ll}
 \textbf{11.} & b_1(1800 r^2 \mu ^2 e^8-2880
   r^2 \mu ^2 e^6-480 r^2 \rho
    e^6+1186 r^2 \mu ^2
   e^4+136 r^2 \rho  e^4-75
   r^2 \mu ^2 e^2+\lambda  \mu
   ^2 e^2 \\
  & +360 r^2 \rho  e^8-1080 r
   \mu ^2 a_0 e^5-360 r \rho
   a_0 e^5+960 r \mu ^2 a_0
   e^3+240 r \rho  a_0
   e^3+\lambda  \rho  e^2-90 r
   \mu ^2 a_0 e \\
  & +1920 r^2 \mu  a_1 e^5-1800
   r^2 \mu  a_1 e^3+180 \mu ^2
   a_0^2 e^2+180 \rho  a_0^2
   e^2-1440 r \mu  a_0 a_1
   e^2+210 r^2 \mu  a_1 e\\
   &+360
   r \mu  a_0 a_1-1080 r \mu ^2 b_1^2
   e^3-360 r \rho  b_1^2
   e^3+540 r^2 a_1^2 e^2+540 r
   \mu ^2 b_1^2 e\\
   &+180 r \rho
   b_1^2 e-180 r^2 a_1^2)=0 \end{array}$ \newline
$\begin{array}{ll}
 \textbf{12.} &b_1(80 r \mu ^3 e^8+48 r \mu
    \rho  e^8-104 r \mu ^3
   e^6-56 r \mu  \rho  e^6+31
   r \mu ^3 e^4+13 r \mu  \rho
    e^4-r \mu ^3 e^2 \\
  & -24 \mu ^3 a_0 e^5-24 \mu
   \rho  a_0 e^5+96 r \mu ^2
   a_1 e^5+12 \mu ^3 a_0
   e^3+12 \mu  \rho  a_0
   e^3-66 r \mu ^2 a_1 e^3+4 r
   \mu ^2 a_1 e \\
  & +32 r \rho  a_1 e^5-18 r
   \rho  a_1 e^3+36 r \mu
   a_1^2 e^2-24 \mu ^2 a_0 a_1
   e^2-24 \rho  a_0 a_1 e^2+6
   \mu ^3 b_1^2 e-6 r \mu
   a_1^2 \\
  & -24 \mu ^3 b_1^2 e^3-24
   \mu  \rho  b_1^2 e^3+6 \mu
   \rho  b_1^2 e)=0 \end{array}$ \newline
$\begin{array}{ll}
 \textbf{13.} & 30 e^2 \left(\mu ^2+\rho
   \right) b_1(60 r \mu ^2 e^6+12 r
   \rho  e^6-56 r \mu ^2 e^4-8
   r \rho  e^4-6 \mu ^2 a_0
   e^3+9 r \mu ^2 e^2 \\
  & -6 \rho  a_0 e^3+64 r \mu
   a_1 e^3-6 \mu ^2 b_1^2 e-6
   \rho  b_1^2 e-24 r \mu  a_1
   e+18 r a_1^2)=0
\end{array}$
$$\,$$

 \noindent\textit{\textbf{Step 3}}. ( This is the more difficult
step ) Solving the previous system for $r$,
 $\mu$, $a_0$, $a_1$, $b_1$ with the aid of \circledR\emph{Mathematica 6.0} we get $b_1=0$, so solutions have
 the form $u=a_0+a_1\sigma(x+\lambda\,t)$. Tables 1,
2, 3 and 4 show the results of calculations.
\begin{center}
\begin{tabular}{|c|c|c|c|}
\hline \,&\,&\,&\\
  $a_0$ & $a_1$ & $\mu$ & $u(x,t);\quad \,\xi=x+\lambda\,t$ \\
           \,&\,&\,&\\ \hline \hline\,&\,&\,&\\
  $\dfrac{1}{60} \left(5 r - \sqrt{5} \sqrt{r^2 - 4 \lambda }\right)$ & $\dfrac{1}{2}$ & $-1$ & $\dfrac{1}{60} \left(5 r - \sqrt{5} \sqrt{r^2 -
              4 \lambda }\right) + \dfrac{r \csc \left(\sqrt{r}
              \,\xi\right)}{2 \left(1 - \csc \left(\sqrt{r} \,\xi\right)\right)}$ \\
           \,&\,&\,&\\  \hline\,&\,&\,&\\
  $\dfrac{1}{60} \left(5 r + \sqrt{5} \sqrt{r^2 - 4 \lambda }\right)$ & $\dfrac{1}{2}$ & $-1$ & $\dfrac{1}{60} \left(5 r + \sqrt{5} \sqrt{r^2 -
              4 \lambda }\right) + \dfrac{r \csc \left(\sqrt{r} \,\xi\right)}{2 \left(1 - \csc \left(\sqrt{r} \,\xi\right)\right)}$
          \\
           \,&\,&\,&\\  \hline\,&\,&\,&\\
  $\dfrac{1}{60} \left(5 r + \sqrt{5} \sqrt{r^2 - 4 \lambda }\right)$ & $-\dfrac{1}{2}$ & $1$ & $\dfrac{1}{60} \left(5 r + \sqrt{5} \sqrt{r^2 -
              4 \lambda }\right) - \dfrac{r \csc \left(\sqrt{r} \,\xi\right)}{2 \left(1+\csc \left(\sqrt{r} \,\xi\right) \right)}$
             \\
           \,&\,&\,&\\  \hline\,&\,&\,&\\
  $\dfrac{\sqrt{\lambda }}{6}$ & $-\dfrac{1}{2}$ & $1$ & $\dfrac{\sqrt{\lambda }}{6} - \dfrac{\sqrt{\lambda } \csc \left(\sqrt[
                4]{4\lambda } \,\xi\right)}{1+\csc \left( \sqrt[
                  4]{4\lambda } \,\xi\right) }$ \\
           \,&\,&\,&\\  \hline
\end{tabular}$$$$
Table 1 : \emph{Solutions for $r>0$, $e=1$ and $\rho=1$. }
\end{center}
\begin{center}
\begin{tabular}{|c|c|c|c|}
  \hline  \,&\,&\,&\\
  $a_0$ & $a_1$ & $\mu$ & $u(x,t);\quad \,\xi=x+\lambda\,t$ \\
           \,&\,&\,&\\ \hline \hline\,&\,&\,&\\
  $\dfrac{1}{60} \left(-5 r - \sqrt{5} \sqrt{r^2 - 4 \lambda }\right)$ & $-\dfrac{i}{2}$ & $-i$ & $\dfrac{1}{60} \left(-5 r - \sqrt{5} \sqrt{r^2 -
              4 \lambda }\right) + \dfrac{1}{2} r \left(1 + \dfrac{1}{\csc \
\left(\sqrt{-r} \,\xi \right) - 1}\right)$ \\
           \,&\,&\,&\\  \hline\,&\,&\,&\\
  $\dfrac{1}{60} \left(- 5 r+\sqrt{5} \sqrt{r^2 - 4 \lambda } \right)$ & $-\dfrac{i}{2}$ & $-i$ & $\dfrac{1}{60} \left(-
        5 r+\sqrt{5} \sqrt{r^2 - 4 \lambda } \right) + \dfrac{1}{2} r \left(1 + \dfrac{1}{\csc \left(\sqrt{-r} \
\,\xi \right) - 1}\right)$ \\
           \,&\,&\,&\\  \hline\,&\,&\,&\\
  $\dfrac{1}{60} \left(- 5 r+\sqrt{5} \sqrt{r^2 - 4 \lambda } \right)$ & $\dfrac{i}{2}$ & $i$ & $\dfrac{1}{60} \left( -
        5 r+\sqrt{5} \sqrt{r^2 - 4 \lambda }\right) + \dfrac{1}{2} r \left(1 - \dfrac{1}{\csc \left(\sqrt{-r} \
\,\xi \right) + 1}\right)$ \\
           \,&\,&\,&\\  \hline\,&\,&\,&\\
  $- \dfrac{5 \sqrt{\lambda }}{6}+\dfrac{\sqrt{\lambda }}{6}$ & $\dfrac{i}{2}$ & $i$ & $\dfrac{\sqrt{\lambda }}{\csc \left(\sqrt[4]{4\lambda } \,\xi \right) +
        1} $ \\
           \,&\,&\,&\\  \hline
\end{tabular}
$$$$ Table 2 : \emph{Solutions for $r<0$, $e=-1$ and $\rho=1$. }
\end{center}
\begin{center}
\begin{tabular}{|c|c|c|c|}
\hline
  $a_0$ & $a_1$ & $\mu$ & $u(x,t);\quad \,\,\xi=x+\lambda\,t$ \\
          \hline\hline\,&\,&\,&\\
  $\dfrac{1}{60} \left(5 r - \sqrt{5} \sqrt{r^2 - 4 \lambda }\right)$ & $\dfrac{1}{2}$ & $-1$ & $\dfrac{1}{60} \left(5 r - \sqrt{5} \sqrt{r^2 -
              4 \lambda }\right) + \dfrac{r \sec \left(\sqrt{r} \,\xi \right)}{2 \
\left(1 - \sec \left(\sqrt{r} \,\xi \right)\right)}$ \\
           \,&\,&\,&\\  \hline\,&\,&\,&\\
  $\dfrac{1}{60} \left(5 r + \sqrt{5} \sqrt{r^2 - 4 \lambda }\right)$ & $\dfrac{1}{2}$ & $-1$ & $\dfrac{1}{60} \left(5 r + \sqrt{5} \sqrt{r^2 -
              4 \lambda }\right) + \dfrac{r \sec \left(\sqrt{r} \,\xi \right)}{2 \
\left(1 - \sec \left(\sqrt{r} \,\xi \right)\right)}$ \\
           \,&\,&\,&\\  \hline\,&\,&\,&\\
  $\dfrac{1}{60} \left(5 r - \sqrt{5} \sqrt{r^2 - 4 \lambda }\right)$ & $-\dfrac{1}{2}$ & $1$ & $\dfrac{1}{60} \left(5 r - \sqrt{5} \sqrt{r^2 -
              4 \lambda }\right) - \dfrac{r \sec \left(\sqrt{r} \,\xi \right)}{2 \
\left(\sec \left(\sqrt{r} \,\xi \right) + 1\right)}$ \\
           \,&\,&\,&\\  \hline\,&\,&\,&\\
  $\dfrac{1}{60} \left(5 r + \sqrt{5} \sqrt{r^2 - 4 \lambda }\right)$ & $-\dfrac{1}{2}$ & $1$ & $\dfrac{1}{60} \left(5 r + \sqrt{5} \sqrt{r^2 -
              4 \lambda }\right) - \dfrac{r \sec \left(\sqrt{r} \,\xi \right)}{2 \
\left(\sec \left(\sqrt{r} \,\xi \right) +
            1\right)}$ \\
           \,&\,&\,&\\  \hline\,&\,&\,&\\
  $-\dfrac{\sqrt{\lambda }}{6}$ & $\dfrac{1}{2}$ & $-1$ & $- \dfrac{\sqrt{\lambda \
}}{6} -\dfrac{1}{2} \sqrt{\lambda }\,\, \text{csch}^2\left(\sqrt[4]{\dfrac{\lambda}{4}}\,\xi\right)$ \\
           \,&\,&\,&\\  \hline\,&\,&\,&\\
  $- \dfrac{\sqrt{\lambda }}{6}$ & $-\dfrac{1}{2}$ & $1$ & $-\dfrac{\sqrt{\lambda }}{6}+\dfrac{\sqrt{\lambda }}{\cosh \left(\sqrt[4]{4\lambda } \,\xi \right) +
        1} $ \\
           \,&\,&\,&\\  \hline\,&\,&\,&\\
  $\dfrac{\sqrt{\lambda \
}}{6}$ & $\dfrac{1}{2}$ & $-1$ & $\dfrac{\sqrt{\lambda
}}{6}+\dfrac{\sqrt{\lambda } \sec \left(\sqrt[
                4]{4\lambda } \,\xi \right)}{1 - \sec \left(\sqrt[
                  4]{4\lambda } \,\xi \right)}$ \\
           \,&\,&\,&\\  \hline
\end{tabular}
$$$$ Table 3 : \emph{Solutions for $e=1$ and $\rho=-1$. }
\end{center}

\begin{center}
\begin{tabular}{|c|c|c|c|}  \hline   \,&\,&\,&\\ $a_0$  &  $a_1$ &  $\mu$  & $u(x,t);\quad
\,\,\xi=x+\lambda\,t$  \\  \,&\,&\,&\\   \hline\hline\,&\,&\,&\\
$\dfrac{1}{60}  \left(-5 r - \sqrt{5} \sqrt{r^2 - 4 \lambda
}\right)$ & $-\dfrac{1}{2}$ & $-1$ & $\dfrac{1}{60}  \left( -5 r -
\sqrt{5} \sqrt{r^2 - 4 \lambda  }\right) - \dfrac{r
\,\text{sech}\left(\sqrt{r} \,\xi
\right)}{2  \left(1  -  \,\text{sech}\left(\sqrt{r}  \,\xi  \right)\right)}$  \\ \,&\,&\,&\\
\hline\,&\,&\,&\\ $\dfrac{1}{60} \left( - 5 r+\sqrt{5} \sqrt{r^2 - 4
\lambda }\right)$ &  $ -\dfrac{1}{2}$ & $-1$ & $\dfrac{1}{60}
\left(- 5 r+\sqrt{5} \sqrt{r^2 - 4 \lambda } \right) - \dfrac{r
\,\text{sech}\left(\sqrt{r}  \,\xi \right)}{2 \left(1 -
\,\text{sech}\left(\sqrt{r} \,\xi \right)\right)}$ \\ \,&\,&\,&\\
\hline\,&\,&\,&\\ $\dfrac{1}{60} \left(-5 r - \sqrt{5} \sqrt{r^2  -
4 \lambda }\right)$  &  $\dfrac{1}{2}$  & $1$  &  $\dfrac{1}{60}
\left(-5 r -  \sqrt{5}  \sqrt{r^2 -  4  \lambda }\right)  + \dfrac{r
\,\text{sech}\left(\sqrt{r}  \,\xi
\right)}{2  \left(\,\text{sech}\left(\sqrt{r}  \,\xi  \right)  +  1\right)}$  \\ \,&\,&\,&\\
\hline\,&\,&\,&\\ $\dfrac{1}{60}  \left(- 5 r+\sqrt{5} \sqrt{r^2  -
4 \lambda }\right)$ & $\dfrac{1}{2}$ & $1$ & $\dfrac{1}{60} \left(-
5 r+\sqrt{5} \sqrt{r^2 - 4 \lambda }\right)  + \dfrac{r
\,\text{sech}\left(\sqrt{r}  \,\xi \right)}{2
\left(\,\text{sech}\left(\sqrt{r}  \,\xi \right) + 1\right)}$  \\
\,&\,&\,&\\  \hline\,&\,&\,&\\  $-\dfrac{\sqrt{\lambda }}{6}$ &  $
-\dfrac{1}{2}$ & $-1$ & $- \dfrac{\sqrt{\lambda
}}{6}-\dfrac{\sqrt{\lambda } \,\text{sech}\left( \sqrt[4]{4 \lambda
} \,\xi  \right)}{1  - \,\text{sech}\left( \sqrt[ 4]{4\lambda  }
\,\xi \right)}
$ \\ \,&\,&\,&\\  \hline\,&\,&\,&\\

$-\dfrac{\sqrt{\lambda  }}{6}$  &  $\dfrac{1}{2}$  &  $1$  & $-
\dfrac{\sqrt{\lambda }}{6}+\dfrac{\sqrt{\lambda  }
\,\text{sech}\left(\sqrt[4]{4\lambda }  \,\xi
\right)}{\,\text{sech}\left(
\sqrt[4]{4\lambda } \,\xi \right) + 1} $ \\
\,&\,&\,&\\  \hline
\end{tabular}
$$$$ Table 4 : \emph{Solutions for $r>0$, $e=-1$ and $\rho=-1$. }
\end{center}

\section{Conclusions}
In this paper, by using the projective Riccati equation method and
the help of symbolic computation system \emph{Mathematica}, we
obtain some exact solutions for the equation (\ref{eq1}).  The
projective Riccati equation method is more complicated than others
methods ( for example, the tanh method ) in the sense that the
algebraic system in many cases demands a lot of time to be solved.
On the other hand, this method allows us to obtain some new exact
solutions.


\bigskip
\begin{thebibliography}{99}

\small
\bibitem{Conte-Musette}{\sc R. Conte \& M. Musette }, \emph{Link betwen solitary waves and projective Riccati
equations}, J. Phys. A Math. \textbf{25} (1992), 5609-5623.

\bibitem{Inc-Mahmut}{\sc E. Inc \& M. Ergüt}, \emph{New Exact Tavelling Wave Solutions for Compound KdV-Burgers Equation in Mathematical
Physics}, Applied Mathematics E-Notes, \textbf{2}(2002), 45-50.

\bibitem{Mei-Zhang-Jiang}{\sc J. Mei, H. Zhang \& D. Jiang},
\emph{New exact solutions for a Reaction-Diffusion equation and a
Quasi-Camassa-Holm Equation}, Applied Mathematics E-Notes,
\textbf{4}(2004), 85-91.

\bibitem{Yan}{\sc Z. Yan}, \emph{The Riccati equation with variable coefficients expansion algorithm to find more exact solutions of nonlinear differential
equation}, MMRC, AMSS, Academis Sinica, Beijing \textbf{22}(2003),
275-284.

\bibitem{Yan}{\sc Z. Yan}, \emph{An improved algebra method and its applications in nonlinear wave euations},
MMRC, AMSS, Academis Sinica, Beijing \textbf{22}(2003), 264-274.

\bibitem{Hereman}{\sc  W.Hereman}, \emph{Symbolic computation of exact solutions
expressible in hyperbolic and elliptic functions for nonlinear
partial differential and diffential-difference equations}, Journal
of Symbolic Computation.

\bibitem{Salas-Gomez}{\sc Alvaro H. Salas \& C. Gomez}, \emph{El mathematica en la b\'usqueda de soluciones exactas para ecuaciones diferenciales parciales lineales y no
lineales}, Primer Simposio Internacional del uso de Tecnolog{\'i}as
en educaci\'on matem\'atica. Universidad Pedag\'ogica
Nacional,Bogotá, Colombia. \textbf{1}(2005).

\bibitem{Salas}{\sc Alvaro H. Salas}, \emph{Some solutions for a type of generalized Sawada-Kotera
equation}, Applied Mathematics and Computation, Elsevier Editorial
System, Netherlands, in press, available on line from july 17 2007
at  http://www.sciencedirect.com

\bibitem{asalas}{\sc C. Gomez, Alvaro H. Salas}, \emph{New exact solutions for the
 combined sinh-cosh-Gordon equation}, Lecturas matem\'aticas,
 volumen especial 2006, Apuntes XV Congreso Nacional de
 Matem\'aticas, agosto de 2005, Bogot\'a, Colombia.

\bibitem{zhao}{XUEQIN ZHAO, HONGYAN ZHI, YAXUAN YU AND HONGQING
ZHANG}, \emph{A new Riccati equation method with symbolic
computation to construct new travelling wave solution of nonlinear
differential equations}, Applied mathematics and Computation 172
(2006) 24-39.
\end{thebibliography}
\end{document}